\documentclass{aa} 

\usepackage{graphicx}
\usepackage{ulem}
\usepackage[T1]{fontenc}

\usepackage[colorlinks=true, allcolors=blue]{hyperref}
\hypersetup{
     colorlinks   = true,
     citecolor    = blue,
     urlcolor = blue
}
\usepackage{txfonts}
\begin{document}

   \title{The Northern Cross Fast Radio Burst project III. The FRB--magnetar connection in a sample of nearby galaxies}

   \author{D. Pelliciari
          \inst{1,2}\fnmsep
          \and
          G. Bernardi\inst{1,3,4}
          \and
          M. Pilia\inst{5}
          \and
          G. Naldi\inst{1}
          \and 
          G. Pupillo\inst{1}
          \and
          M. Trudu\inst{5,6}
          \and
          A. Addis\inst{7}
          \and
          G. Bianchi\inst{1}
          \and
          C.~Bortolotti\inst{1} 
          \and
            D.~Dallacasa\inst{1,2} 
          \and
            R.~Lulli\inst{1}
          \and
            A.~Maccaferri\inst{1}
          \and
            A.~Magro\inst{8}
          \and
            A.~Mattana\inst{1}
          \and
            F.~Perini\inst{1}
          \and
            M.~Roma\inst{1}
          \and
            M.~Schiaffino\inst{1}
          \and
          G. Setti\inst{1,2}
          \and
            M.~Tavani\inst{9,10} 
          \and
          F. Verrecchia\inst{11,12}
          \and
          C. Casentini\inst{9}}

   \institute{INAF-Istituto di Radio Astronomia (IRA), via Piero Gobetti 101, Bologna, Italy\\
              \email{davide.pelliciari@inaf.it}
         \and
             Dipartimento di Fisica e Astronomia, Universit\'{a} di Bologna, Via Gobetti 93/2, 40129 Bologna, Italy
         \and
             South African Radio Astronomy Observatory, Black River Park, 2 Fir Street, Observatory, Cape Town, 7925, South Africa
         \and
             Department of Physics and Electronics, Rhodes University, PO Box 94, Makhanda, 6140, South Africa
         \and
             INAF-Osservatorio Astronomico di Cagliari, via della Scienza 5, I-09047, Selargius (CA), Italy
        \and
            Università degli Studi di Cagliari, Dipartimento di Fisica, SP Monserrato-Sestu km 0.7, I-09042 Monserrato (CA), Italy
        \and
            INAF-Osservatorio di Astrofisica e Scienza dello Spazio di Bologna, Via Piero Gobetti 93/3, 40129, Bologna, Italy
        \and
            Institute of Space Sciences and Astronomy (ISSA), University of Malta, Msida, MSD 2080, Malta
        \and
            INAF/IAPS, via del Fosso del Cavaliere 100, I-00133 Roma (RM), Italy
        \and
            Università degli Studi di Roma "Tor Vergata", via della Ricerca Scientifica 1, I-00133 Roma (RM), Italy
        \and
            SSDC/ASI, via del Politecnico snc, I-00133 Roma (RM), Italy
        \and
            INAF-Osservatorio Astronomico di Roma, via Frascati 33, 00078 Monte Porzio Catone (RM), Italy}

   \date{XXX-XXX-XXX}

 
  \abstract
   {Fast radio bursts (FRBs) are millisecond radio transients observed at cosmological distances. The nature of their progenitors is still a matter of debate, although magnetars are invoked by most models. The proposed FRB--magnetar connection was strengthened by the discovery of an FRB-like event from the Galactic magnetar SGR~J1935+2154.
   }
   {In this work we aim to investigate how prevalent magnetars such as SGR~J1935+2154 are within FRB progenitors.
   }
   {To this end, we carried out an FRB search in a sample of seven nearby (< 12 Mpc) galaxies with the Northern Cross Radio Telescope for a total of 692 h.
   }
   {We detected one 1.8 ms burst in the direction of M101 with a fluence of $58 \pm 5$ Jy ms. Its dispersion measure of 303 pc cm$^{-3}$ places it most likely beyond M101. Considering that no significant detection comes indisputably from the selected galaxies, we place a 38 yr$^{-1}$ upper limit on the total burst rate (i.e. including the whole sample) at the 95\% confidence level. This upper limit constrains the event rate per magnetar to $\lambda_{\rm mag} < 0.42$~magnetar$^{-1}$ yr$^{-1}$ or, if combined with literature observations of a similar sample of nearby galaxies, it yields a joint constraint of $\lambda_{\rm mag} < 0.25$ magnetar$^{-1}$ yr$^{-1}$. We also provide the first constraints on the expected rate of FRBs hypothetically originating from ultra-luminous X-ray (ULX) sources, since some of the galaxies observed during our observational campaign host confirmed ULXs. We obtain $< 13$ yr$^{-1}$ per ULX for the total sample of galaxies observed.}
   {Our results indicate that bursts with energies $E > 10^{34}$ erg from magnetars such as SGR~J1935+2154 appear more rarely compared to previous observations and further disfavour them as unique progenitors for the cosmological FRB population. This provides support to the idea that there is a greater contribution from a population of more exotic magnetars not born via core-collapsed supernovae.}

   
   \titlerunning{FRB-magnetar connection in nearby galaxies}

   \maketitle

\section{Introduction}\label{sec:Intro}

Fast radio bursts (FRBs) are millisecond-long, highly dispersed radio signals with exceptionally high brightness temperatures ($10^{32} - 10^{35}$~K) of extragalactic origin \citep{Thornton13, CordesChatterjee19, petroff19, petroff21,Pilia22}. Since their discovery \citep{Lorimer2007}, about $600$ distinct FRB sources have been observed \citep{CHIME_Cat}, with a host galaxy association for about 25 of them \citep{Tendulkar17, Bannister19, Ravi19_loc, Prochaska19, Maquart19, Marcote20, Bhardwaj21, Ravi21, Niu21, Nimmo21, Ryder22, Ravi22, Bhandari22_nonrep, Driessen23, Sharma23}. Moreover, 50 FRBs were found to be repeating \citep{CHIME23}, with recent statistical evidence indicating that repeaters may represent a distinct class of sources \citep{Hashimoto20, Pleunis21, Zhong22, Guo22, CHIME23}. Up to now, only two repeaters, FRB~20180916B and FRB~20121102, have shown periodic activity, with periods of $\sim 16.33 \pm 0.12$ \citep{Pleunis21} and $\sim 161 \pm 5$ days \citep{Cruces21}, respectively. 

Of the progenitor candidates, magnetars -- neutron stars (NSs) powered by the decay of their strong ($10^{14} - 10^{16}$ G) magnetic field \citep{DuncanThompson, Rea11, Turolla15, Kaspi17} -- are the most widely considered.
The FRB--magnetar connection was strengthened by the detection of FRB 20200428, the first Galactic FRB-like event discovered \citep{CHIME20b, Bochenek20a}; it was observed simultaneously with an X-ray burst \citep{Mereghetti20, Ridnaia20, Li20, Tavani20}, emitted by the soft gamma repeater J1935+2154 (SGR1935 --hereafter), one of the most active known magnetars \citep{Stamatikos2014, Lien2014, Cummings2014, Kozlova16, Younes17a}. 
The reported isotropic-equivalent energy emitted at radio wavelengths by this burst, $E_{\rm SGR} \simeq 2 \times 10^{34}$~erg \citep{Bochenek20a, Margalit2020}, lies between those of the energetic pulsar giant radio pulses \citep[see e.g.,][]{Kuzmin07} and  extragalactic FRBs. Further observations revealed fainter bursts from SGR1935 \citep{Zhang20, Burgay2020, Good20, Rodin20, Kirsten21, Dong22, Maan22, Huang22, Pearlman22}.

Since magnetars are connected to a young stellar population, they are expected to be found in regions of star formation. While this has been observed for some FRBs \citep{Chatterjee17, Marcote17, Bassa17, Tendulkar17, Ravi19_loc, Marcote20, Bhardwaj21, Niu21, Piro21, Tendulkar21, Nimmo21}, others have been found in galaxies with low star formation rates (SFRs), mainly in their outskirts \citep{Heintz20, Mannings21, Bhandari22}. Notably, the repeating FRB 20200120E was recently found in a globular cluster in the nearby galaxy (NG) M81 \citep{KirstenM81}. It is also theoretically possible to find young magnetised NSs in globular clusters, formed either via the accretion-induced collapse of a white dwarf \citep[WD;][]{Tauris13, Wang20} or the collapse of a compact binary system induced by a WD-WD, WD-NS or NS-NS merger \citep{Giacomazzo13, Schwab16, Zhong20}.

\citet{CHIME20b} observed a sample of 15 NGs within 12~Mpc with SFRs higher than that of the Milky Way (MW) in order to search for SGR1935-like FRBs--(i.e. bursts with energy $E_0 > 4 \times 10^{34}$~erg). The advantage of targeting NGs over the MW is that the whole population of magnetars is simultaneously observed, therefore increasing the probability of detecting a burst with respect to individual magnetars. No bursts were observed, allowing them to place a constraint on the burst rate of SGR1935-like events $0.007 < \lambda_{\rm mag} < 0.4$~yr$^{-1}$~magnetar$^{-1}$, where the lower limit is set by the detection of the SGR1935 burst itself \citep{CHIME20b}.

In this work we present long monitoring observations of a sample of NGs taken with the Northern Cross (NC) Radio Telescope in the search for FRBs with the aim to study the FRB--magnetar connection, following \cite{CHIME20b}. We note that some of the galaxies observed within our observational campaign host confirmed ultra--luminous X-ray (ULX) sources, accreting binary systems with X-ray luminosities $L_X \geq 10^{39}$ erg s$^{-1}$, which is significantly higher than any X-ray luminosity emitted by stellar processes \citep{Fabbiano85, Kaaret17}. 

The paper is structured as follows: Sect.~\ref{sec:obs} describes the observations and the sample of NGs observed, Sect.~\ref{sec:upper_limits} presents constraints on the FRB burst rate of magnetars (Sect. \ref{sec: upp_lim_mag}) and ULXs (Section \ref{sec: upp_lim_ulx}) determined using these observations. Finally, the implications that our constraints have on the connection between FRBs and their progenitors, along with our conclusions, are discussed in Sect~\ref{sec:conclusions}. 

\section{An FRB search in a sample of nearby galaxies}
\subsection{Sample description and observations}
\label{sec:obs}
\begin{table*}
	\centering
        \caption{Properties of the observed NG sample. Columns list: coordinates, distances ($D$), SFRs, the dispersion measure contribution from the MW interstellar medium (DM$_{\rm ISM}$) computed as the maximum between the YM16 \citep{YMW16} and NE2001 \citep{CordesLazio02,CordesLazio03} models, and the total exposure time spent on source. SFRs are estimated from H$\alpha$ luminosities for all the galaxies but M82 and IC~342, for which infrared luminosities were used. References for distances are: [1] \citet{McConnachie05}; [2] \citet{Kara04}; [3] \citet{Dalcanton09}; [4] \citet{Shappe11}; [5] \citet{Anand18}; [6] \citet{Newman01}; and [7] \citet{Hoyt19}. References for SFRs are: [8] \citet{Rahmani16}; [9] \citet{Gao&Solomon04}; [10] \citet{Forster03}; [11] \citet{Kennicutt08}.}
	\begin{tabular}{lcccccc}
		\hline
		  & R.A. (J2000) & Dec (J2000) & D [Mpc] & SFR [$M_\odot\ \rm yr^{-1}$] & DM$_{\rm ISM}$ [pc~cm$^{-3}$] & $\rm T$ [hr]\\
		\hline
		\hline
		M31 & $00^{\rm h} 42^{\rm m} 44.3^{\rm s}$ & $+41^\circ 16' 07.5''$ & $0.79 \pm 0.03$ [1] & $0.35$ [8] & $142$ & $51$\\
		IC342 & $03^{\rm h}46^{\rm m}48.5^{\rm s}$ & $+68^\circ 05' 46.0''$ & $3.3 \pm 0.3$ [2] & $2.8$ [9] & $178$ & $102$ \\
		M82 & $09^{\rm h} 55^{\rm m} 52.4^{\rm s}$ & $+69^\circ 40'46.9''$ & $3.53 \pm 0.04$ [3] & $13$ [10] & $41.2$ & $184$\\
		M101 & $14^{\rm h} 03^{\rm m} 12.6^{\rm s}$ & $+54^\circ 20'55.5''$ & $6.4 \pm 0.5$ [4] & $2.9$ [11] & $30.9$ & $96$\\
		NGC6946 & $20^{\rm h} 34^{\rm m} 52.3^{\rm s}$ & $+60^\circ 09'13.2''$ & $7.7 \pm 0.3$ [5] & $4.3$ [11] & $145.8$ & $115$\\
		M106 & $12^{\rm h} 18^{\rm m} 57.6^{\rm s}$ & $+47^\circ 18'13.4''$ & $7.8 \pm 0.6$ [6] & $2.8$ [11] & $25.8$ & $84$\\
		M66 & $11^{\rm h} 20^{\rm m} 15.0^{\rm s}$ & $+12^\circ 59'28.6''$ & $11.1 \pm 0.4$ [7] & $2.7$ [11] & $31.1$ & $63$\\
		\hline
	\end{tabular}
	\label{table_gal}
\end{table*}
\begin{figure}
    \centering
	\includegraphics[width=1.0\columnwidth]{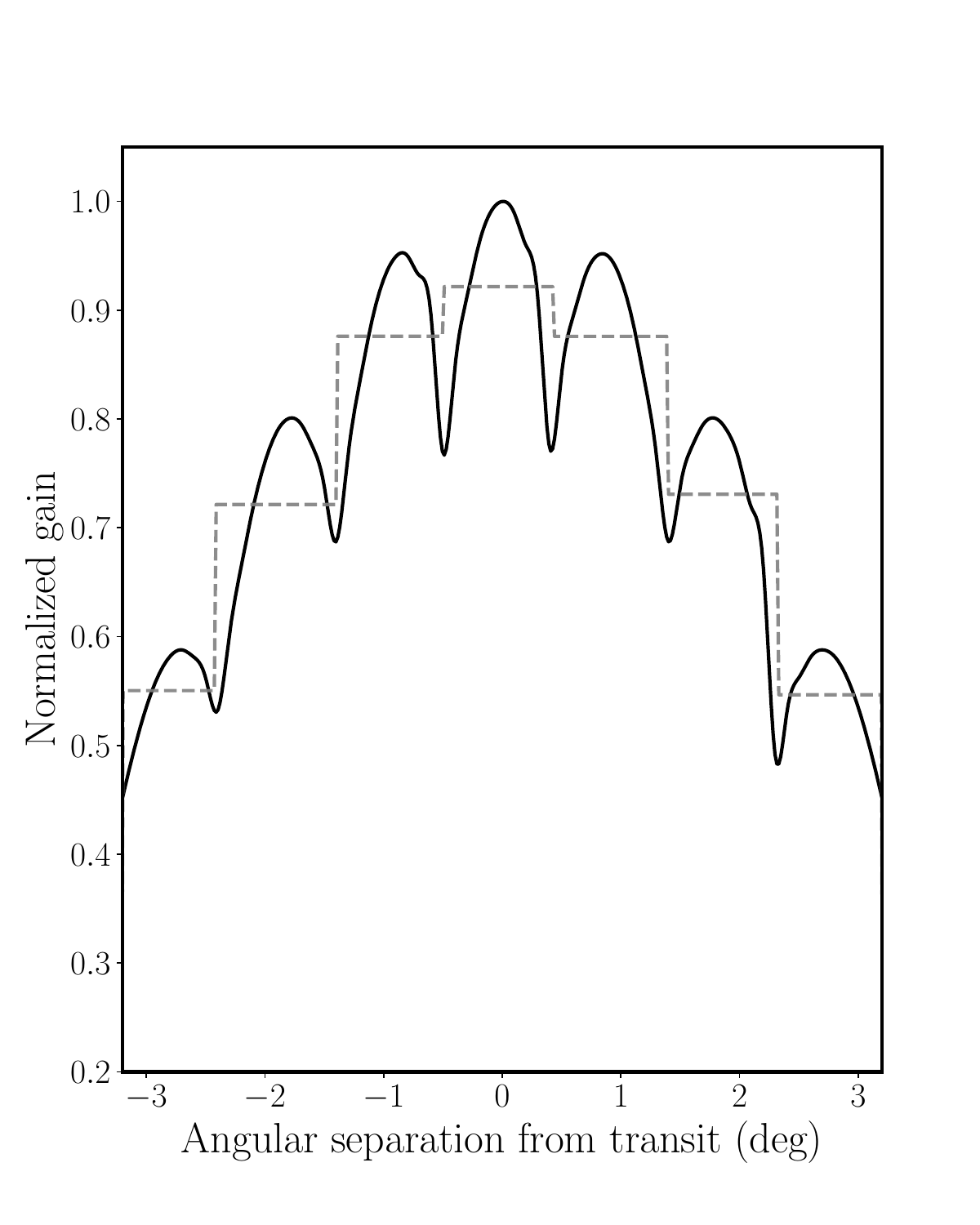}
    \caption{Normalized NC beam-forming response (solid line) as the source transits across the telescope field of view. Dashed lines represent the average value of the telescope beam response for each delay bin.}
    \label{fig:beam}
\end{figure}
Our sample consists of seven galaxies within a maximum distance of 12~Mpc, whose characteristics are summarized in Table~\ref{table_gal}. We selected them based on their reported high SFRs, since magnetars such as SGR1935, born through core-collapse supernovae (CCSNe), trace young star formation sites. Apart from M31 and IC 342, all the galaxies were already included in the sample observed by \cite{CHIME20b}. We included M31 for its proximity and similarity to the MW. As already mentioned, among the galaxies observed, M82, M101, IC 342 and NGC 6946 host confirmed ULXs. They are M82 X-1 \citep{Ptak99}, M101 X-1 \citep{Stetson98}, IC 342 X-1, IC 342 X-2 \citep{Rana2015} and NGC 6946 X-1 \citep{Fabbiano87, Roberts03}. These objects are currently believed to be the result of super-Eddington accretion onto a stellar mass black hole \citep[BH; e.g.][]{Liu13}, or a NS \citep[][and references therein]{Pintore20}.

We monitored each galaxy daily during its transit through the telescope's primary beam using eight cylinders of the north-south arm of the array \citep[we refer the reader to][for details of the system]{Locatelli20}. Due to limitations of the current acquisition software, a source cannot be tracked continuously as it moves across the telescope's field of view; therefore, we employed the 'shift and track' strategy used in \cite{Trudu22}, where seven discrete delay values are approximately equally spaced in angular size to cover the field of view. The resulting telescope beam pattern is shown in Fig.~\ref{fig:beam}, where the seven peaks corresponding to the seven beam-forming delays are clearly visible. Such a beam pattern implies that the sensitivity to the source varies by up to 40\% as it transits through the primary beam. 
As the variation in the beam response within each delay bin is small ($\approx 10\%$), in the following analysis we assume it to be constant and equal to its average value (dashed line in Fig.~\ref{fig:beam}).
The energy detection threshold, $E_{\rm min}$, for a burst observed in the delay bin $j$ from the galaxy $i$ can be written as
\begin{equation}\label{eq:NC_lum_limit}
    E_{{\rm min},i,j} = 4 \pi D_i^2 \,  \Delta \nu \, \frac{F}{A_j},
\end{equation}
where $D_i$ is the source distance, $\Delta \nu = 16$~MHz the observing bandwidth, $A_j$ the average beam gain in the $j$-th delay bin and $F = 38$~Jy~ms the NC fluence threshold corresponding to the peak of the central beam assuming a $10\sigma$ detection threshold \citep{Trudu22}. In this work we assume 1~ms as the reference burst duration.

Observations started on December 26, 2021, and ended on August 21, 2022, for a total of 692~hr. Data were stored to disk with a time resolution of 138.24~$\mu$s and a frequency channel width of 14.468~kHz. We calibrated our data using interferometric observations of Cas~A \citep[see][for details on the calibration procedure]{Locatelli20}, carried out approximately at the beginning, halfway and at the end of the observational campaign.  Receiver phases and amplitudes remained fairly constant over week-long timescales; nevertheless, our observing campaign was interspersed with monthly observations of the pulsar B0329+54 and the repeating FRB~20180916B \citep{chime18}. Both sources served as calibration tests for the stability of our system and, at the same time, FRB~20180916B was observed within its window of expected activity \citep{chime18,PastorMarazuela21,Pleunis21,Trudu22} for the purpose of obtaining multiwavelength observations \citep{Pilia2020,Tavani2020, Trudu22}. 

We regularly detected single pulses from B0329+54 at its nominal dispersion measure (DM), $\sim 26.8$~pc~cm$^{-3}$ \citep{Hassal12}, and detected two new bursts from FRB~20180916B, in addition to those presented in \citet{Trudu22}, in agreement with the expected repetition rate.
\begin{figure*}
    \centering
	\includegraphics[width=0.9\textwidth]{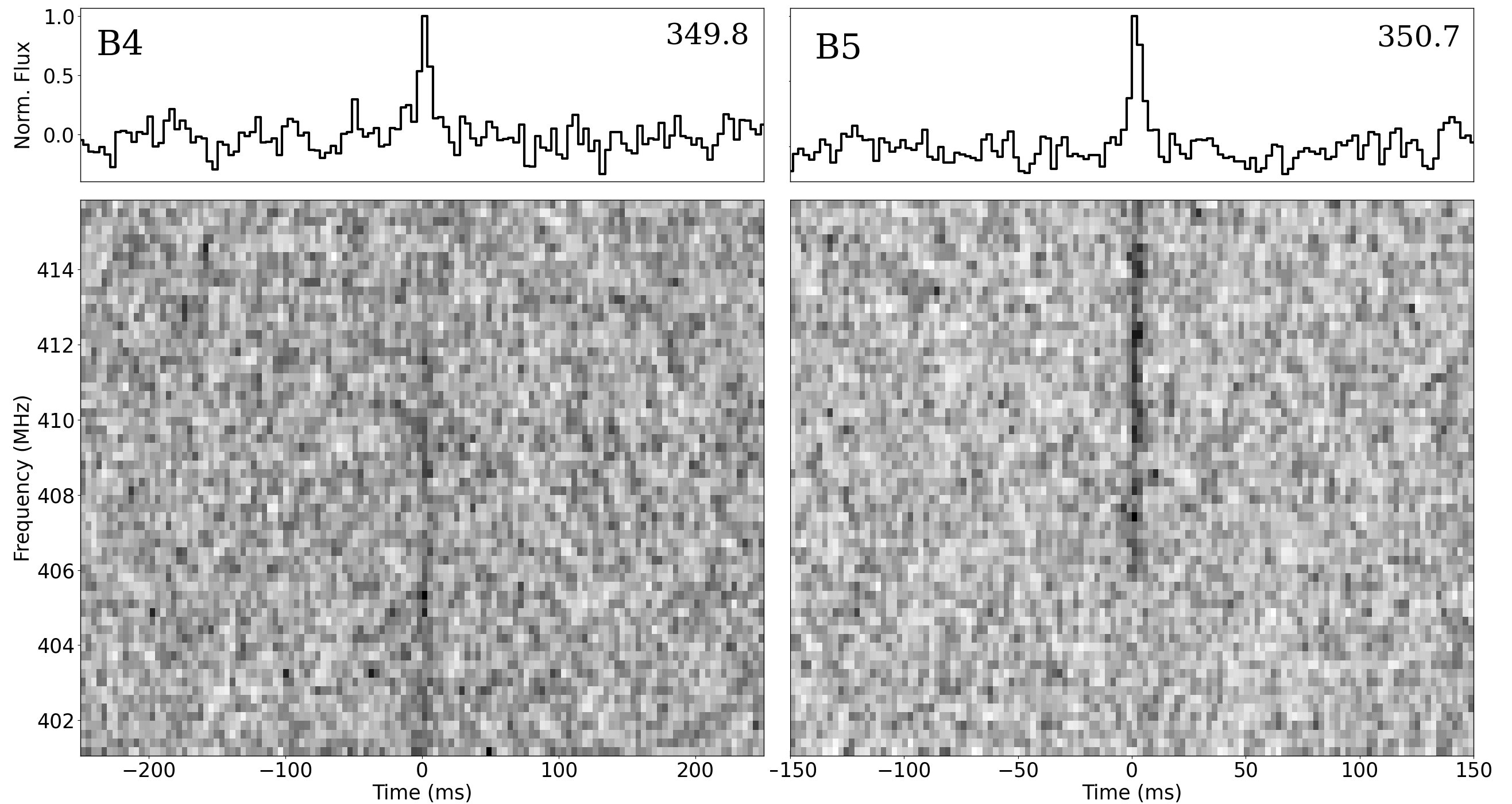}
    \caption{Bursts from FRB~20180916B observed on April $30^{\rm th}$ 2022 (left panel) and August $9^{\rm th}$ 2022 (right panel), respectively. Bottom panels show the dynamic spectra, while the top panels show the frequency averaged profiles.The best-fit DMs (in pc~cm$^{-3}$) at which the bursts were de-dispersed are reported in the top right corner of each plot. Data were down-sampled to have 64 frequency channels, each $0.25$~MHz wide, and time bins with $2.2$~ms width for better display.}
    \label{fig:R3_bursts}
\end{figure*}
\begin{table}
	\caption{Properties of B4 and B5 bursts from FRB 20180916B. We report, from the top row to the bottom, the barycentric time of arrival (TOA) expressed as the modified julian day (MJD), the signal to noise ratio (S/N), the fit-optimized DM, the full width at half maximum (FWHM) duration, the flux density and the fluence of the bursts.}	
	\begin{tabular}{lcc} 
        \hline
		 Parameter  & B4 & B5 \\
		\hline
		\hline
		TOA (MJD) & 59699.52603591 & 59800.25184782\\
		S/N & $15$ & $20$ \\
		DM (pc cm$^{-3}$) & $349.8 \pm 0.1$ & $350.7 \pm 0.1$\\
		$\Delta t$ (ms) & $6.35 \pm 0.1$ & $5.7 \pm 0.3$ \\
		Flux density (Jy) & $15 \pm 1$ & $19 \pm 1$\\
		Fluence (Jy ms) & $96 \pm 6$ & $108 \pm 5$\\
		\hline
	\end{tabular}
	\label{tab:R3_bursts_prop}
\end{table}

Figure~\ref{fig:R3_bursts} shows their de-dispersed waterfall plots, labelled B4 and B5 to follow the \cite{Trudu22} nomenclature, and their best-fit values are listed in Table~\ref{tab:R3_bursts_prop}. These detections provide evidence for calibration stability across the whole campaign.

The search for FRBs in our galaxy sample was performed using the {\sc SPANDAK} pipeline \citep{Gajjar18}, which flags radio frequency interference (RFI) through {\sc rfifind} \citep{Ransom2002} and searches for single pulses with {\sc Heimdall} \citep{bbb+12}. We used the same search setup described in \citet{Trudu22}, within the $0 < {\rm DM} < 1000$~pc~cm$^{-3}$ range, with a signal-to-noise ratio (S/N) greater than six and a boxcar duration shorter than 35~ms in order to balance the search time and our scientific goals whilst remaining consistent with the observed width distribution of FRBs \citep{petroff21, CHIME_Cat}.

\subsection{FRB detection from the direction of M101}

\begin{figure}
    \centering
	\includegraphics[width=1.0\columnwidth]{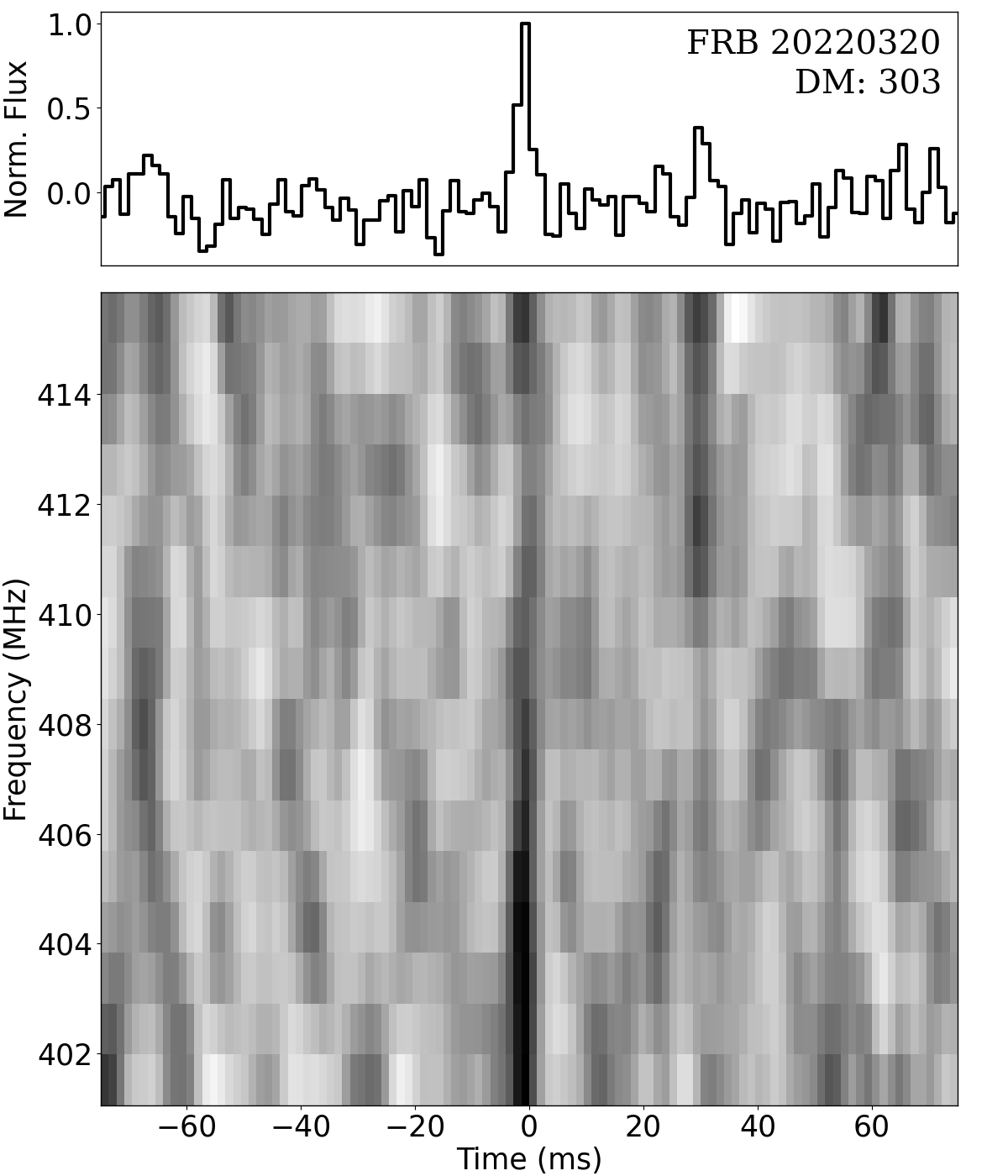}
    \caption{De-dispersed profile of FRB~20220320. The top panel shows the frequency-averaged time series, and the bottom panel shows its dynamic spectrum, coherently de-dispersed at DM~$= 303$~pc~cm$^{-3}$.}
    \label{frb220320}
\end{figure}

We found no candidates with a S/N~$> 10$ throughout the whole campaign and $\sim 100$ candidates with $ 6 \leq {\rm S/N} \leq 10$ that were further visually inspected and double-checked. Most candidates were discarded due to the presence of RFI contamination and noise properties that showed deviations from a theoretical Gaussian distribution. The only candidate that conservatively passed the selection, detected with a $\rm S/N \sim 11$, is shown in Fig.~\ref{frb220320}.  

We assessed the likelihood of a candidate signal by computing the number of false candidates with $\rm S/N$ higher than the detection threshold, expected because of noise outliers. The number of false positive events is a function of the S/N and can be computed as
\citep{CordesMcLaughlin03}
\begin{equation}
N_{\rm false}(> {\rm S/N}) = 2 N_{\rm DM} \, \frac{\Delta T}{{\rm d}t} \, P(> \rm S/N),
\end{equation}
where $N_{\rm DM}$ is the number of DM trials considered in the FRB search, set internally in {\sc Heimdall}, $\Delta T$ the total duration of the observations, ${\rm d}t$ the time sample duration, and
\begin{equation}
P(>{\rm S/N}) = \frac{2}{\sqrt{\pi}} \int_0^x e^{-z^2} {\rm d} z \equiv \frac{1}{2} \Biggl[1-{\rm erf}\Biggl(\frac{{\rm S/N}}{\sqrt{2}}\Biggr)\Biggr]
\end{equation}
the probability that an event is due to a noise statistical fluctuation. In our case, $N_{\rm DM} \sim 3000$, $\Delta T = 692$~h, and ${\rm d}t \sim 138\ \mu$s, resulting in $N_{\rm false} \sim 10^{-9}$ for $\rm S/N \geq 10$. Hence, we consider our candidate to be a genuine FRB.

The burst, hereafter FRB 20220320, was detected on 2022 March 30 UT = 01:14:02.40 (Barycentric time of arrival, $\infty$ MHz), in the direction of M101, and its properties are listed in
Table~\ref{tab:fbr_properties}.
\begin{table}
	\caption{Observational properties of FRB 20220320. The uncertainty on the FRB position corresponds to the beam full width at half maximum.}	
	\begin{tabular}{lc} 
        \hline
		Parameter & Value \\
		\hline
		\hline
		R.A. (J2000, deg) & $211(3)$ \\
		Dec. (J2000, deg) & $54.4(5)$ \\
		T.o.A & 2022-03-20 01:14:02.40 (UT)\\
		S/N & $11$\\
		DM (pc cm$^{-3}$)& $303 \pm 2$\\
		$\Delta t\ (\rm ms) $ & $1.8 \pm 0.3$ \\
		Flux density (Jy) & $32 \pm 3$ \\
		Fluence (Jy ms) & $58 \pm 5$ \\
		\hline
	\end{tabular}
	\label{tab:fbr_properties}
\end{table}
We measured a DM~$ = 303 \pm 2$~pc~cm$^{-3}$, which disfavours a local origin. The DM contribution along the line of sight of M101 due to the interstellar medium (ISM) is $\rm DM_{\rm ISM} \simeq 23 - 31$~pc~cm$^{-3}$, according to the NE2001 \citep{CordesLazio02,CordesLazio03} and the YMW16 \citep{YMW16} electron density models, while the contribution from the Galactic halo is $\sim 50$~pc~cm$^{-3}$ \citep{Agarwal19, Maquart19, Yamasaki, Lemos22}. Finally, the intergalactic medium (considering $z \simeq 0.001$) contributes approximately $\sim 2$~pc~cm$^{-3}$ \citep{Maquart19}, although this value, for low redshift galaxies, depends on the line of sight \citep{Li19}. These contributions, if coming from M101, would imply a moderately high $\rm DM_{\rm host}$, $\simeq 220$~pc~cm$^{-3}$. In our case, the almost face-on inclination of M101 seems to disfavour a large $\rm DM_{\rm host}$ for M101 \citep{Xu2015}. Although \citet{James22}, by analysing 16 FRBs from Australian Square Kilometre Array Pathfinder (ASKAP) observations, found $\langle \rm DM_{\rm host} \rangle \sim 186 \pm 50$ pc cm$^{-3}$, which would be compatible with an origin from M101, other works point towards lower values for the host DM contribution \citep{Niino20, Zhang20}. This is further evidence that an association with M101 is unlikely.

If we consider the burst to be originating farther away, assuming a more conservative value $\rm DM_{\rm host} \simeq 100$~pc~cm$^{-3}$, the burst would be placed at a maximum redshift of $z \sim 0.18$, that is to say, at a luminosity distance of $\sim 870$~Mpc. For the distance estimation we considered a seven-year Wilkinson Microwave Anisotropy Probe (WMAP7) cosmology \citep{WMAP7} to be consistent with \citet{Maquart19}. Given this maximum redshift, its spectral luminosity would be at most $6.3 \times 10^{33}$~erg~s$^{-1}$~Hz$^{-1}$, well within the range of extragalactic FRB spectral luminosities \citep{Bochenek20a, CHIME20b, Luo2020}. 

\section{Upper limits on the FRB repetition rate from our observations}
\label{sec:upper_limits}

In this section we show how the observations conducted in our NG campaign allowed us to extract important upper limits on the FRB burst rate. In particular, we developed a simple model to calculate the expected rate of FRB events from the whole sample of galaxies listed in Table \ref{table_gal}, considering as a first case a single population of SGR1935-like magnetars as FRB progenitors. By taking $\lambda_{\rm mag}$, the average burst rate per magnetar, and $\gamma$, the power-law energy distribution slope for magnetar bursts, as free parameters, we discuss how the upper limits on the rate from the whole galaxy sample translate into upper limits on $\lambda_{\rm mag}$ and $\gamma$. We also discuss how our observations coul constrain the FRB event rate if FRBs originated from ULXs. Indeed, accretion-based mechanisms for the FRB engine have been proposed \citep{Waxman17, Katz17, Katz2020, Sridhar21, Deng21}. 

\subsection{SGR1935-like magnetars}\label{sec: upp_lim_mag}
We started by computing the total burst rate from our galaxy sample, considering a single population of magnetars similar to SGR1935. This means that we restricted our analysis to magnetars that have radio efficiencies $\eta \sim 10^{-5}$, the ratio of energy radiated in the radio and X-ray bands by SGR1935 \citep{Mereghetti20,Tavani20,Ridnaia20,CHIME20b,Bochenek20a}, and burst energetics $E \geq E_0$, where $E_0 = 2 \times 10^{34}$ erg is the isotropic radio energy released by FRB 20200428 \citep{Bochenek20a, Margalit2020, CHIME20b}. This energy follows by considering $d = 7$ kpc to be the distance to SGR1935 \citep{Margalit2020}.

Following \citet{CHIME20b}, we assumed that the magnetar burst rate $\mathcal{R}$ from a given galaxy is proportional to its SFR. We computed the number of SGR1935-like magnetars residing in a certain galaxy by scaling for $N_{\rm mag}/\rm SFR_{\rm MW}$, with $N_{\rm mag} = 29$ \citep{Olausen14, Kaspi17} the number of Galactic magnetars and SFR$_{\rm MW} = 1.65 \pm 0.19$ M$_{\odot}$ yr$^{-1}$ the Galactic SFR \citep{Licquia15}.

With these prescriptions, the rate expected from the galaxy $i$ of our sample can be expressed as
\begin{eqnarray}\label{eq:R_init}
\begin{aligned}
    \mathcal{R} (\lambda_{\rm mag}, \gamma; > E_{{\rm min}, i,j}) &= N_{\rm mag} \, \frac{\rm SFR_{i}}{\rm SFR_{\rm MW}} \times \\ &\times \int_{E_{{\rm min}, i,j}}^{E_{\rm max}} K_0 \, \left( \frac{E}{E_0} \right)^{-\gamma} \Theta [E - E_0] \, dE,
    \label{eq:predict_upper_limits_bin}
\end{aligned}
\end{eqnarray}
where $E_{{\rm min}, i,j}$ is the minimum burst energy detectable from the galaxy $i$ in the delay bin $j$ (Eq.~\ref{eq:NC_lum_limit}). The integral is the burst rate energy function, which we assumed to follow a power law with index $\gamma$. We note that the Heaviside function, $\Theta$, restricts the case to SGR1935-like bursts (i.e. with energy $E \geq E_0$). We considered the maximum energy, $E_{\rm max}$, to be
\begin{equation}
    E_{\rm max} = \eta \, E_{\rm mag},
\label{eq:mag_energy}    
\end{equation}
with $\eta \sim 10^{-5}$ and $E_{\rm mag}$ the total magnetic energy reservoir for a magnetar with magnetic field $B$ \citep{Margalit2020}:
\begin{equation}
    E_{\rm mag} = 3 \times 10^{49} \, \left ( \frac{B}{10^{16} \, {\rm G}} \right )^2 \, \, {\rm erg}.
\label{eq:mag_field_energy}
\end{equation}
We considered $B = 2 \times 10^{14}\ \rm G$, the magnetic field strength of SGR1935 \citep{Israel16}, and verified that a higher value for $B$ does not appreciably affect our results.

The burst rate normalization, $K_0$, can be expressed as a function of the burst rate per magnetar $\lambda_{\rm mag}$: 
\begin{equation}
    \lambda_{\rm mag} = \int_{E_0}^{E_{\rm max}} K_0 \, \left( \frac{E}{E_0} \right)^{-\gamma} \, dE,
\label{eq:lamba_mag_norm}
\end{equation}
yielding:
\begin{equation}
    K_0 = \lambda_{\rm mag} \frac{(1-\gamma) \, E_0^{-\gamma}}{E_{\rm max}^{-\gamma + 1} - E_0^{-\gamma + 1}}.
\label{eq:k0}    
\end{equation}
\begin{figure}
    \centering
	\includegraphics[width=1.0\columnwidth]{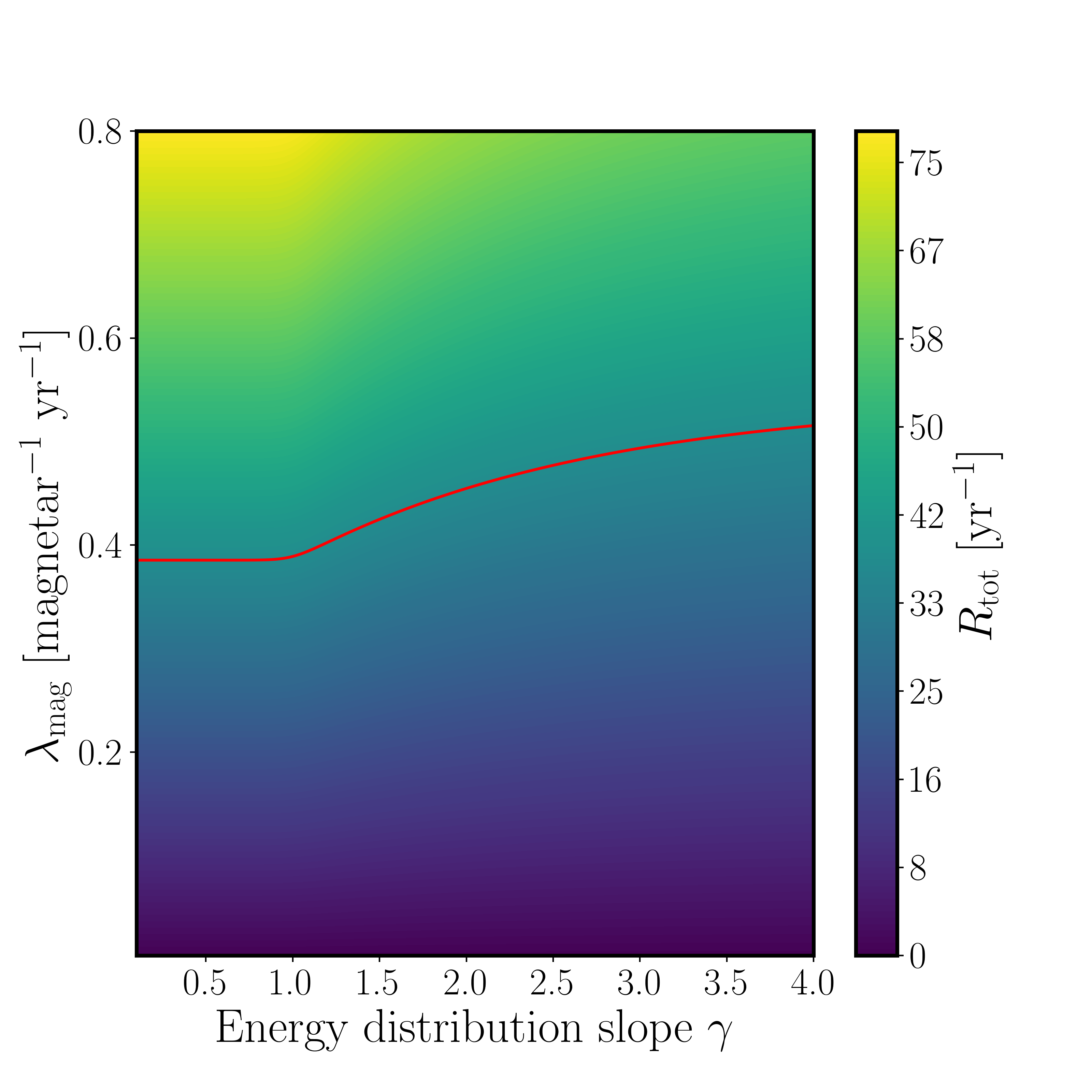}
    \caption{Expected burst rate (colour-map) as a function of the energy distribution slope, $\gamma$, and the SGR1935-like burst rate per magnetar, $\lambda_{\rm mag}$, from the whole observing campaign of NGs. The red line is the 95\% CL upper limit obtained via our observations.}
    \label{fig:burst_rate_tot}
\end{figure}
Equation~\ref{eq:predict_upper_limits_bin} can therefore be re-written as
\begin{eqnarray}
    \mathcal{R} (\lambda_{\rm mag}, \gamma; > E_{{\rm min},i,j}) = N_{\rm mag} \, \lambda_{\rm mag} \, \frac{\rm SFR_{i}}{\rm SFR_{\rm MW}} \, \frac{E_{\rm max}^{-\gamma + 1} - E_{{\rm min},i,j}^{-\gamma + 1}}{E_{\rm max}^{-\gamma + 1} - E_0^{-\gamma + 1}} \ .
\label{eq:predict_upper_limits_final_bin}
\end{eqnarray}
We could then compute the total expected burst rate from our NG sample as the sum of the rate for each galaxy in each delay bin, weighted by their corresponding integration time:
\begin{equation}\label{eq:predicted_tot}
    \mathcal{R}_{\rm tot} (\lambda_{\rm mag}, \gamma) =  \frac{\sum_i^N \sum_j^{N_b} \mathcal{R} (\lambda_{\rm mag}, \gamma; > E_{{\rm min},i,j}) \, T_{i,j}}{\sum_i^N  \sum_j^{N_b} T_{i,j}},
\end{equation}
where $N_b = 7$ is the number of delay bins and $T_{i,j}$ is the total integration time of the $i$-th galaxy in the delay bin $j$. The total expected burst rate is mildly dependent upon the slope of the energy distribution and much more significantly dependent upon the burst rate per magnetar as can be visually assessed in Fig.~\ref{fig:burst_rate_tot}. Our observations led to a 95\% confidence level \citep[CL;][]{Gehrels86} upper limit on the total burst rate of $\mathcal{R}_{\rm tot} < 38$~yr$^{-1}$. By imposing this upper limit, we obtain a 95\% CL constraint on the burst rate per magnetar that varies between $\lambda_{\rm mag} < 0.38$~magnetar$^{-1}$~yr$^{-1}$ for $\gamma \leq 1$ and $\lambda_{\rm mag} < 0.47$~magnetar$^{-1}$~yr$^{-1}$ for $\gamma \simeq 3$. After marginalizing over the energy distribution slope $\gamma$, we derive $\lambda_{\rm mag} < 0.42$~magnetar$^{-1}$~yr$^{-1}$. 

Previous works derived constraints for the magnetar burst rate. In particular, \citet{CHIME20b} obtained $0.007 < \lambda_{\rm mag} < 0.4$ magnetar$^{-1}$ yr$^{-1}$ at the 95$\%$ CL from the detection of FRB 20200428 and the non-detection from a sample of 15 NGs. However, the authors adopted slightly different values than us for the distances and SFRs of the galaxies that are common to our sample (see their Extended Data Table 1), and they considered a lower $\rm SFR_{\rm MW}$ of $1\ M_\odot$ yr$^{-1}$. If we use their values in our estimate, we obtain $\lambda_{\rm mag} < 0.3$ magnetar$^{-1}$ yr$^{-1}$. By combining our observations with \cite{CHIME20b}, we can obtain an even further improved upper limit on $\lambda_{\rm mag}$. Assuming the distance and SFR values presented in Table~\ref{table_gal} for both samples, we recomputed the total rate expected from the joint observations (Eqs.~\ref{eq:predict_upper_limits_final_bin} and \ref{eq:predicted_tot}), which has to be compared with the upper limit obtained from the total observing time ($\sim 1370 \ \rm h$). As a result, we obtain $\lambda_{\rm mag} < 0.25$ magnetar$^{-1}$ yr$^{-1}$ at the 95$\%$ CL, the most stringent upper limit on the burst rate per magnetar to date. In Sect. \ref{sec:conclusions} we discuss this result in the framework of the FRB--magnetar connection. 

\subsection{Ultra-luminous X-ray sources}\label{sec: upp_lim_ulx}
An alternative formation channel to the more standard magnetar models predicts that FRBs can be produced by short-lived relativistic flares from super-Eddington accreting NSs and BHs \citep{Sridhar21}. The isotropic energy emitted by this kind of FRB is postulated to lie in the range $10^{34}$ erg - $10^{45}$ erg \citep{Sridhar21}, which implies that these sources would have been detected with the NC sensitivity. Therefore, the observing time spent on these galaxies is useful for placing an upper limit on the rate of FRBs produced by these accreting objects.

We followed the approach described in Sect.~\ref{sec:upper_limits} and expressed the rate expected from the population of ULXs, $\mathcal{R}_{u,i}$, in the galaxy $i$ of our sample: 
\begin{equation}
    \mathcal{R}_{u,i} (\lambda_u) = N_{u,i} \, \lambda_u, 
\end{equation}
where $N_{u,i}$ is the number of ULX sources expected in the galaxy $i$ and $\lambda_u$ is the average burst rate per ULX for energies in the $10^{34} < E < 10^{45}$~erg range. The number of ULX sources can be expressed as a function of the SFR 
\citep{Kovlakas20}:
\begin{equation}\label{eq: nulx}
    N_{u,i} = 0.51 \, \frac{{\rm SFR}_i}{{\rm M}_\odot {\, \rm yr}^{-1}}.
\end{equation}
We expect from Eq. \eqref{eq: nulx} a total of $\sim 14$ ULXs in the galaxies we observed. The total expected burst rate, $\mathcal{R}_{u, {\rm tot}}$, from our NG sample then becomes:
\begin{equation}
        \mathcal{R}_{u, {\rm tot}} (\lambda_u) = \frac{\sum_i^N \sum_j^{N_b} \mathcal{R}_{u,i} (\lambda_u) \, T_{i,j}}{\sum_i^N  \sum_j^{N_b} T_{i,j}},
\end{equation}
The observed upper limit at the 95$\%$ CL on the total burst rate is $\simeq 38$ yr$^{-1}$, the same as determined in Sect.~\ref{sec: upp_lim_mag}, since it depends only on the total observing time. Therefore, by imposing $\mathcal{R}_{u, {\rm tot}} < 38$ yr$^{-1}$, we obtain $\lambda_u < 13$ yr$^{-1}$ for the average burst rate per ULX. This is the first upper limit on the FRB repetition rate hypothetically coming from ULXs. We note that this upper limit is two to three orders of magnitude lower than the  reported repetition rate of most active repeaters from FRB 20121102A and FRB 20180916B, $r \sim 10^3$ yr$^{-1}$ \citep{Margalit2020}. We also find this contrast by computing the repetition rate for confirmed ULXs only, that is to say, without estimating the population of ULXs from Eq. \eqref{eq: nulx} and only considering already discovered ULXs in the galaxies we observed. We report the upper limits obtained for each case in Table \ref{tab: ulx} and discuss the possible implications of these results in the following section.

\begin{table}
	\centering	
        \caption{Upper limits for the FRB repetition rate (last column) coming from  confirmed ULXs. In the second column we report the number of ULXs present in a given galaxy. The last row considers the total expected ULX ensamble derived from Eq. \eqref{eq: nulx}.}
	\begin{tabular}{lccc} 
		\hline
		  ULX & $N_{\rm u}$ & $\lambda_{\rm u}$ [yr$^{-1}$]\\
            \hline
		\hline
            IC 342 X-1, IC 342 X-2 & $2$ & $< 128$ \\
            M82 X-2 & $1$ & $< 140$ \\
            M101 X-1 & $1$ & $< 273$ \\
            NGC 6946 X-1 & $1$ & $< 227$\\
            All & $14$ & $< 13$\\
            \hline
	\end{tabular}
	\label{tab: ulx}
\end{table}

\section{Discussion and conclusions}
\label{sec:conclusions}
In this paper we present an FRB search in a sample of seven NGs taken with the NC Radio Telescope. The campaign was 692~h long and yielded the detection of a 58~Jy~ms, 1.8~ms long burst with a DM~=~303~pc~cm$^{-3}$, observed in the direction of M101, although most likely coming from a more distant source. Therefore, we consider that no detections came from the monitored galaxies. We used this result to investigate the connection between FRBs and magnetars, by computing the total burst rate, $\mathcal{R}_{\rm tot}$, expected from our galaxy sample, assuming, as unique FRB progenitors, magnetars such as SGR J1935+2154, and obtaining $\mathcal{R}_{\rm tot} < 38 $~yr$^{-1}$ at the 95$\%$ CL. We considered $\lambda_{\rm mag}$, the average burst rate per magnetar, and $\gamma$, the slope of the burst energy distribution, as free parameters for our SFR-based model. In addition, we derived the average FRB rate per ULX from the same observations of NGs. Among the target galaxies of our observational campaign, some host confirmed ULXs, from which, at least theoretically, an FRB would have obtained enough energy \citep{Sridhar21} to be detected by the NC.

Since the detection of the Galactic FRB-like signal, it has been tempting to claim that magnetars such as SGR1935 (i.e. magnetars similar in energetics and formed through CCSNe) represent the entire cosmological population of FRBs. Although the implied volumetric rate from the detection of FRB 20200428 is consistent with the faint energy of the cosmological rate density \citep{LuBenKum22}, SGR1935-like magnetars cannot explain the high repetition rate of active repeaters \citep{Margalit2020}. Moreover, FRB 20200428 is about one order of magnitude fainter than the average FRB \citep{Bochenek20a, CHIME20b, Luo2020}, although giant magnetar flares may be bright enough to fill the gap \citep{Margalit2020}. The discrepancy in the repetition rate, along with the discovery of the M81 repeater localized in a globular cluster \citep{Bhardwaj21, Kirsten21}, led to consider other, more exotic NS formation channels \citep{Kremer21, Kremer22}, and question the presence of a single population of magnetars as FRB progenitors.

Our observations do not constrain the burst energy slope well, although they somewhat disfavour flat slopes ($\gamma < 1$) over steeper ones ($\gamma > 1$). The average burst rate per magnetar is, instead, constrained to be $\lambda_{\rm mag} <0.42$~magnetar$^{-1}$~yr$^{-1}$.
This upper limit halves the range for the magnetar burst rate implied by the detection from the Survey for Transient Astronomical Radio Emission 2 \citep[STARE2;][]{Bochenek20a},  $0.0036 < \lambda_{\rm mag} < 0.8$~magnetar$^{-1}$~yr$^{-1}$, consistently with the results from Canadian Hydrogen Intensity Mapping Experiment's (CHIME) NG observations, $0.007 < \lambda_{\rm mag} < 0.4$~magnetar$^{-1}$~yr$^{-1}$ \citep{CHIME20b}. We also show how the upper limit lowers to $\lambda_{\rm mag} < 0.25$ magnetar$^{-1}$ yr$^{-1}$ if we consider our NG observations combined with the monitoring reported in \citet{CHIME20b}.

Considering the STARE2 detection, \citet{Margalit2020} already pointed out that the cosmological FRB rate, including repeating sources, can be explained only by adding a second magnetar population that is younger and with a stronger magnetic field with respect to the SGR1935-like ones \citep{Margalit19,Blanchard16}. This second population includes more exotic magnetars, not born through the usual supernova core collapse, but through a much rarer formation channel. Our constraints imply that the burst rate per magnetar of SGR-like events is approximately a factor of two smaller than previously reported, implying rarer events. From the perspective of a two-population model, our results imply that rare magnetars should be more prominent than considered earlier in order to compensate for the smaller burst rate from SGR1935-like magnetars. Moreover, the detection of FRB 20200120E in a globular cluster \citep{Bhardwaj21a, Kirsten22glob} has similar implications \citep{LuBenKum22}, namely that the formation of magnetars is possible inside globular clusters through compact object mergers, accretion-induced collapse, or a WD-merger-induced collapse \citep{Kirsten22glob, Kremer21, Kremer22}. However, the nanosecond structures observed in some bursts from FRB 20200120E could be explained in terms of a recycled millisecond pulsar origin \citep{Majid21, Kremer21}. Hence, it is not clear whether M81-like FRBs could represent this needed population of rarer FRB progenitors.

Among other exotic but prominent progenitor models, we find accretion-based mechanisms \citep{Waxman17, Katz17, Katz2020, Sridhar21, Deng21}. In particular, the recent model of \citet{Sridhar21} is able to explain both the energetics and the chromaticity behaviour seen in FRB 20180916B \citep{PastorMarazuela21}. Estimating the total number of ULXs present in our sample of galaxies to be 14, we have constrained the average rate of FRB events per ULX to $\lambda_u < 13$ yr$^{-1}$ at the $95\%$ CL. We have also provided individual upper limits for each monitored ULX. These initial estimates show a discrepancy of a few orders of magnitude when compared to the repetition rates of the most active repeaters \citep{Margalit2020}. As already pointed by \citet{Sridhar21}, a strongly magnetised object in the compact pair could be necessary to power cosmological FRB luminosities, making these events much rarer and more difficult to detect. In conclusion, our limits on the burst rate disfavour both magnetars and ULXs as progenitors of very active repeating sources such as FRB 20180916B and FRB 20121102A.

\begin{acknowledgements}
      We thank the anonymous referee for the useful comments, which helped us improving the quality of the manuscript. The reported data were collected during the phase of the INAF scientific exploitation with the Northern Cross radio telescope.
\end{acknowledgements}

\bibliography{biblio}{}

\bibliographystyle{aasjournal}

\end{document}